\documentclass[prl,twocolumn]{revtex4}

\usepackage{graphicx}
\usepackage{dcolumn}
\usepackage{epsfig}
\usepackage{latexsym}
\usepackage{amsfonts}
\usepackage{amssymb}
\begin{document}
\bibliographystyle{apsrev}
\title{ Self-diffusion of Rod-like Viruses Through Smectic Layers}

\author{M. Paul Lettinga$^1$ and Eric Grelet$^2$}
\affiliation{$^1$ IFF, Institut Weiche Materie, Forschungszentrum
J\"{u}lich, D-52425 J\"{u}lich, Germany \\
$^2$ Centre de Recherche Paul Pascal, CNRS-Universit\'e Bordeaux
1, 115 Avenue Schweitzer, 33600 Pessac, France }

\date{\today}
\begin{abstract}

We report the direct visualization at the scale of {\it single}
particles of mass transport between smectic layers, also called
permeation, in a suspension of rod-like viruses. Self-diffusion
takes place preferentially in the direction normal to the smectic
layers, and occurs by {\it quasi-quantized} steps of one rod
length. The diffusion rate corresponds with the rate calculated
from the diffusion in the nematic state with a lamellar periodic
ordering potential that is obtained experimentally.

\end{abstract}


\maketitle

Since the pioneering work of Onsager on the entropy driven phase
transition to a liquid crystalline state \cite{Onsager}, the
structure and the phase behavior of complex fluids containing
anisotropic particles with hard core interactions has been a
subject of considerable interest, both theoretically
\cite{rodsTheo} and experimentally \cite{rodsExp}. Understanding
of the particle mobility in the different liquid crystalline
phases is more recent \cite{simulations}. In experiments various
methods have been applied to obtain the ensemble averaged
self-diffusion coefficients in thermotropic \cite{NMRthermo} and
amphiphilic \cite{NMRlyo} liquid crystals, block copolymer
\cite{Fredrickson96} and colloidal systems \cite{FRAP}. Only a few
studies have been done where dynamical phenomena are probed at the
scale of a {\it{single}} anisotropic particle: the Brownian motion
of an isolated colloidal ellipsoid in confined geometry
\cite{ellipsoid} and the self-diffusion in a nematic phase formed
by rod-like viruses \cite{EPLpavlik} represent two recent
examples. In the latter case, the diffusion parallel ($D_\| $) and
perpendicular ($D_\bot $) to the average rod orientation (the
director) has been measured, showing an increase of the ratio
$D_\|/D_\bot $ with particle concentration. Knowledge of the
dynamics at the single particle level is fundamental for
understanding the physics of mesophases with spatial order like
the smectic (lamellar) phase 
of rod-like particles. In this mesophase the particle density is
periodic in one dimension parallel to the long axis of the rods,
while the interparticle correlations perpendicular to this axis are
short-ranged (fluid-like order). For parallel diffusion to take
place, the rods need to jump  between adjacent smectic layers,
overcoming an energy barrier related to the smectic order parameter
\cite{deGennes}. This process of interlayer diffusion, or
{\it{permeation}}, was first predicted by Helfrich \cite{Helfrich}.
In this Letter, we use video fluorescence microscopy to monitor
the dynamics of individual labeled colloidal rods in the
background of a smectic mesophase formed by identical but
unlabeled rods. In this way we directly observe {\it{permeation}}
of single rods in adjacent layers. As in the nematic phase,
self-diffusion in a smectic phase is anisotropic: the diffusion
through the smectic layers is shown here to be much faster than
the diffusion within each liquid-like layer, i.e. $D_\|/D_\bot \gg
1$, in contrast to thermotropic systems. Moreover, since the
individual rod positions within the layer are monitored, the
potential barrier for permeation is straightly determined for the
first time. The permeation can then be described in terms of
Brownian particles diffusing in a one-dimensional periodic
symmetric potential.

\begin{figure}
\includegraphics[width=.4\textwidth]{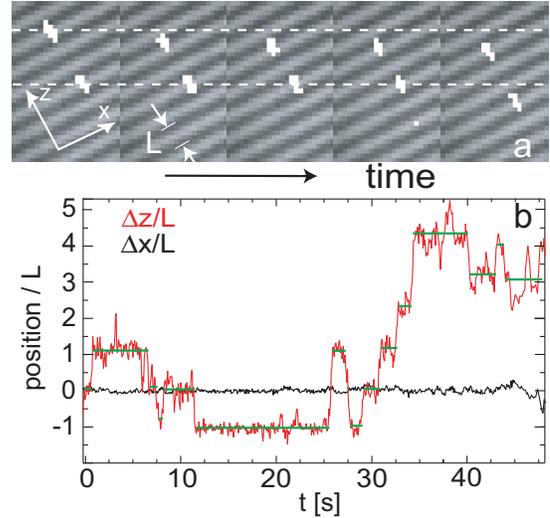}
\caption{\label{jumpingrods} (a) Time sequence of an overlay of
fluorescence and DIC images showing labeled particles jumping
between adjacent smectic layers ($\Delta t =0.071 s$ is the time
between two frames). The 
layer spacing is $L\simeq0.9~\mu m$. (b) Displacement of a given
particle in the direction parallel (red) and perpendicular (black)
to the director. The green lines indicate the residence time, i.e.
the time for which one particle stays in a given layer. }
\end{figure}

The system of rods used in this work consists of filamentous
bacteriophages {\it fd}, which are semi-rigid polyelectrolytes
with a contour length of 
0.88~$\mu m$, a diameter of
6.6~nm, and a persistence length of
2.2~$\mu m$ \cite{protocol}. Suspensions of {\it{fd}} rods in
aqueous solution form several lyotropic liquid crystalline phases,
in particular the chiral nematic (cholesteric) phase and the smectic
phase \cite{Revuefd}. The existence of a smectic phase in
suspensions of hard rods is an evidence of the high monodispersity
and therefore of the model system character of such filamentous
viruses \cite{SmPoly,KRP-PRE2007}. The colloidal scale of the
{\it{fd}} bacteriophage facilitates the imaging of individual rods
by fluorescence microscopy, as well as smectic layers by
differential interference contrast (DIC) microscopy \cite{Revuefd}.
Fig. \ref{jumpingrods}(a) shows a sequence of images of a single
region \cite{protocolVideo} where both techniques are combined. A
comparison of the images shows that some rods jump between two
layers while others remain within a given layer. The trajectory of
one of the rods is plotted in Fig. \ref{jumpingrods}(b) in the
direction parallel (z) and perpendicular (x) to the director. This
figure summarizes the key observation of this Letter: the diffusion
throughout the smectic layers takes place in {\it quasi-quantized
steps} of one rod length i.e. the mass transport between the layers
is a discontinuous process. Moreover, it shows that the diffusion
within the smectic player is extremely slow \cite{note}.

\begin{figure}
\includegraphics[width=.4\textwidth]{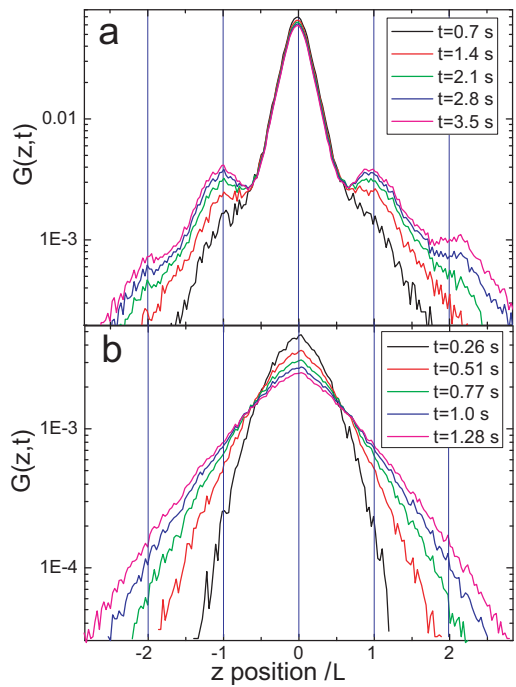}
\caption{\label{vanHove} Probability density function in space at
different times for two ionic strengths: 20~mM (a) and 110~mM (b).
The functions are normalized to one, the z-axis is scaled by the
smectic layer thickness $L$.
}
\end{figure}

The ``hopping-type'' diffusion is mainly the consequence of the
underlying ordering potential of the smectic phase and the
vacancies available in adjacent layers.
A phenomenological expression for permeation has been derived by
coupling the displacement of a segment of a smectic layer $u$ to
the compressibility modulus $\tilde{B}$ via the permeation
parameter $\lambda_b $ \cite{deGennes}:

\begin{eqnarray}\label{Eq_permeation1}
\frac{\partial u}{\partial t}=\lambda_b \tilde{B} \frac{\partial^2
u}{\partial z^2}.
\end{eqnarray}
On a single-particle level, the fundamental solution of this
diffusion equation is the self-van Hove function \cite{VanHove},
which is the probability for a displacement $z$ during a time $t$:

\begin{eqnarray}\label{Eq_vanHove}
G(z,t)=\frac{1}{N}\sum_{i=1}^{N} \delta
[z+z_{i}(0\frac{}{})-z_{i}(t)].
\end{eqnarray}
Since single particles are experimentally identified, the self-van
Hove function can be directly obtained from the histogram of
particle positions after a time $t$, as plotted in Fig.
\ref{vanHove} for low (I~=~20~mM) and high ionic (I~=~110~mM)
strengths. For a  fluid made of Brownian particles, a smooth
gaussian distribution that smears out over time is expected for
the self-van Hove function. However at low ionic strength,
$G(z,t)$ shows distinct peaks exactly  at integer multiples of the
particle length (and therefore of the layer thickness, see Fig.
\ref{vanHove}(a)) as expected from visual observation (Fig.
\ref{jumpingrods}). At high ionic strength the curves are smoother
(Fig. \ref{vanHove}(b)), but in all cases the experimental
self-van Hove function is not gaussian at any time. This implies
that the permeation parameter $\lambda_b$ in Eq.
\ref{Eq_permeation1} is a function of position $z$, due to the
energy landscape imposed by the smectic layers.

\begin{figure}
\includegraphics[width=.4\textwidth]{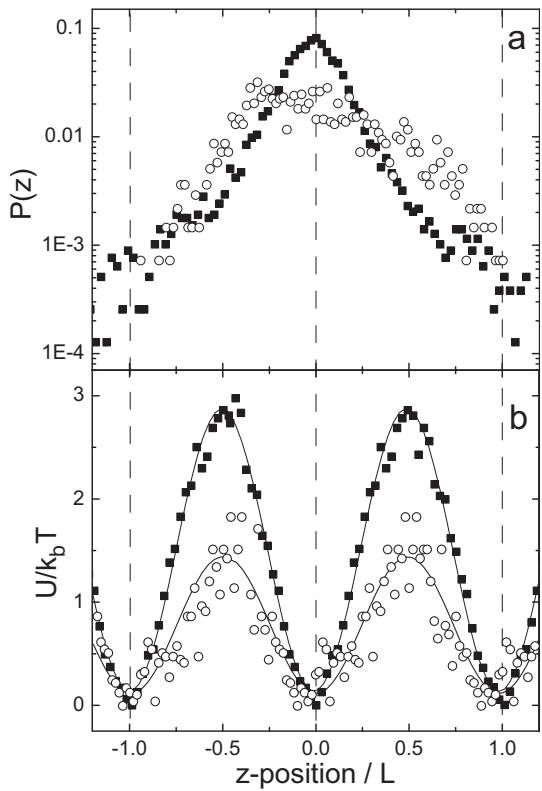}
\caption{\label{SmecPotential} (a) Histogram of time averaged
particle positions parallel to the director {\it{within}} the
smectic layer  at 20~mM ($\blacksquare$) and 110~mM ($\circ$). The
histogram is normalized by the total number of positions. (b)
Resulting effective mean ordering potential in the z-direction
obtained by applying the Boltzmann factor. The solid lines are a
fit to a sinusoidal potential. }
\end{figure}

The energy landscape can be determined experimentally from the
distribution of particle positions with respect to the middle of a
layer parallel to the director. To this end, time windows are
selected where the particle remains for ten frames or more within
the same layer. The distribution of particles within a single
layer is then obtained by addition of all particle positions
relative to the average position of particles for all selected
time windows. The resulting distributions are plotted in Fig.
\ref{SmecPotential}(a) for the two ionic strengths. To obtain the
total particle distribution for the full smectic phase, the
distributions of particles in a single layer (Fig.
\ref{SmecPotential}(a)) is added periodically to itself at all
integer numbers of layer spacing $L$ (
Fig. \ref{jumpingrods}(a)). The smectic ordering potential is then
deduced from the Boltzmann factor $P(z)\sim
e^{-U_{layer}(z)/k_BT}$ for the probability of finding a particle
at position $z$, as shown in Fig. \ref{SmecPotential}(b). Both
potentials can be best fitted with a sinusoidal $U_{layer}(z)=U_0
\sin(2\pi z/L)$, giving an amplitude of $U_0=1.36\;k_BT$ at low
ionic strength and $U_0=0.66\;k_BT$ at high ionic strength. The
difference between the two amplitudes explains the fact that for
I~=~20~mM the self-van Hove function exhibits discrete peaks,
while for I~=~110~mM the potential barrier is small enough to
exhibit a monotonic behavior of the probability density function.
The reason for the more pronounced potential at low ionic strength
might be that electrostatic interactions between rods are more
long-ranged, i.e., particles are more strongly correlated so that
it is more difficult to create a void between them. The fact that
the potential can be fitted by a sinusoidal is remarkable by
itself. Indeed, the use of such a potential is very common due to
its simplicity \cite{SmTheo}, but this ordering potential has
never been directly observed until now. Moreover the height of the
potential, i.e. the smectic order parameter, can be directly
obtained.

\begin{figure}
\includegraphics[width=.4\textwidth]{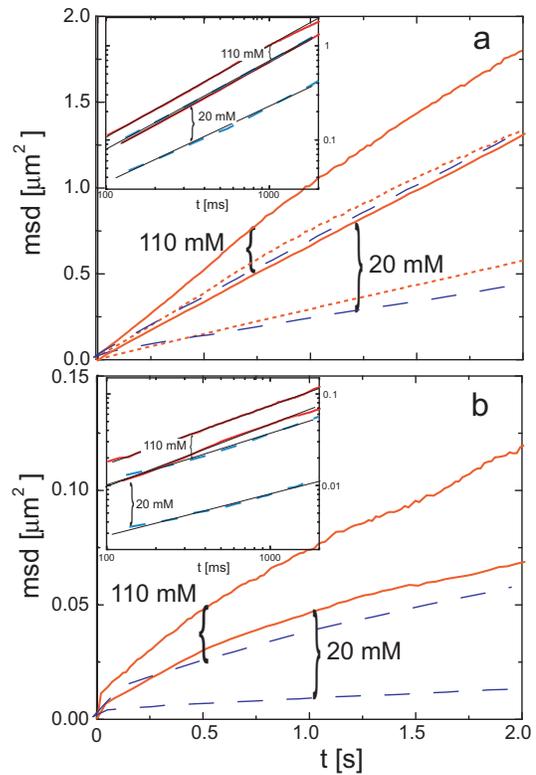}
\caption{\label{MSD} (a) Mean square displacement parallel to the
director for the two indicated ionic strengths in the nematic
phase (red line), in the smectic phase (blue dashed line), and in
the nematic phase considering the oscillatory potential (red
dotted line). (b) Mean square displacement perpendicular to the
director (same convention as above). The insets show the same data
in a log-log scale, yielding the degree of subdiffusion from the
linear regression (black line).}
\end{figure}

The overall mean square displacement (MSD) of rods parallel and
perpendicular to the director of the smectic and nematic phase is
plotted in Fig. \ref{MSD} for both ionic strengths. The time
evolution of the MSD given by $\langle \Delta r^2(t)\rangle \sim
t^{\gamma}$ provides the diffusion exponent $\gamma$: $\gamma <1$ is
characteristic of a {\it{sub}}diffusive behavior, while $\gamma
>1$ is referred to as {\it{super}}diffusion. The parallel motion is
close to be diffusive ($\gamma\simeq1$) in both the nematic
($\gamma=0.97$) and smectic ($\gamma=0.94$) phases for I~=~110~mM
and in the nematic phase for I~=~20~mM ($\gamma=0.95$). Only the
parallel motion in the smectic phase for low ionic strength, i.e.
where the discrete peaks in the self-van Hove function are
observed, is significantly subdiffusive: $\gamma=0.81$. The
perpendicular motion is in all cases strongly subdiffusive: for
I~=~110~mM, $\gamma$ reduces from 0.63 before to 0.56 after the
nematic-smectic (N-Sm) transition and for I~=~20~mM it reduces
from 0.68 to 0.46. Anomalous subdiffusive behavior has often been
observed in systems where diffusion takes place by steps, e.g. in
case of release from a surrounding cage \cite{cage}. This ``cage
escape'' might be at the origin of the observed subdiffusive
behavior for both parallel and perpendicular diffusion. For
parallel diffusion the cage is formed by the energy barrier
imposed by the smectic layers, as shown by smaller $\gamma$ for
higher ordering potential. Perpendicular diffusion at high volume
fractions is only possible through a reptation-like motion along
the long axis to escape the local excluded volume, as observed for
polymers for which typically $\gamma=0.5$ \cite{McLeish02}. This
excluded volume is huge, even for thin rods at high orientational
order, due to the large rod aspect ratio of $\approx 130$. In
addition, perpendicular diffusion in the smectic phase is hindered
due to the ordering potential, which couples this diffusion to the
permeation and which thus explains the decrease of $\gamma$ from
the nematic to the smectic phases. For subdiffusive systems, a
non-Gaussian distribution of the probability density functions has
been observed as in Fig. \ref{vanHove} \cite{cage}, even though
these two features are not {\it{a priori}} correlated. Note also
that boundary effects might influence the probability density
\cite{wall}.

The anisotropy in the total diffusion, $D_\|/D_\bot$, which is about
20 in the nematic phase \cite{EPLpavlik}, increases in the smectic
phase as a result of the pronounced subdiffusivity of the
perpendicular motion (decrease of $\gamma$). These observations show
an opposite trend as compared to thermotropic liquid crystals
\cite{simulations,NMRthermo}, where usually $D_\|/D_\bot$ evolves
from being larger than one at temperatures close to the N-Sm
transition temperature to being smaller than one at lower
temperatures \cite{sm}. Therefore the diffusion in the smectic phase
can be effectively considered as a one-dimensional diffusion of a
Brownian particle in a periodic potential in the high friction
limit. A general expression for such a diffusion process is given by
\cite{Festa78}:

\begin{eqnarray}\label{Eq_Msd}
{D_\|} =\frac{D_0}{\langle e^{-U_{layer}(z)/k_BT}\rangle \langle
e^{U_{layer}(z)/k_BT}\rangle}.
\end{eqnarray}
The brackets indicate averaging over one period of the ordering
potential. The diffusion coefficient in the smectic phase can then
be calculated taking $D_0$ as the diffusion coefficient in the
nematic phase close to the N-Sm transition, and using $U_{layer}$
as obtained from the fit of the potentials plotted in Fig.
\ref{SmecPotential}: the diffusion coefficient decreases by a
factor 0.84 at I~=~110~mM and by a factor 0.44 at I~=~20~mM.
Indeed the MSD in the smectic phase is obtained from the MSD in
the nematic phase, using these factors for both ionic strengths
(see Fig. \ref{MSD}), although at I~=~20~mM some deviation appears
due to the subdiffusivity in the MSD. Thus, we have shown how the
mobility of rods decreases after the N-Sm transition, contrary to
the isotropic-nematic transition where the global mobility
increases due to entropic gain \cite{Onsager,EPLpavlik}. It seems
therefore to indicate that {\it fd} virus suspensions do not
behave as a system of rigid hard rods for high concentration in
agreement with a recent work \cite{KRP-PRE2007}. Moreover, the
very slow diffusion within the layers suggests that the smectic
phase of semi-flexible colloidal rods consists of layers of
glass-like, rather than fluid-like, particles.

In conclusion, we have for the first time visualized the process of
permeation in the smectic phase at the scale of single particles for
a system of charged rods. This allowed us to give a full and
coherent description of the diffusion process without any
assumptions on the system. The diffusion is strongly anisotropic in
the direction normal to the smectic layers and quasi-discontinuous
due to the presence of the layers. The parallel diffusion rate
complies with the rate in the nematic phase, taking into account the
ordering potential, which is obtained directly from our
measurements.

We thank Jan Dhont for fruitful discussions. This project was
supported by the European network of excellence SoftComp. MPL
acknowledges also the SFB TR6 for financial support.

\clearpage


\begin{references}
\bibitem{protocolVideo} The time resolution of the used CCD camera (Hamamatsu C9100) was between 10 and 120
ms, depending on the region of interest.
\bibitem{note} To quantify this hopping-type diffusion, the jumps have been
identified with a home-written algorithm based on the criterion
that the first N positions in time should deviate in one direction
from the previous average, and that the distance between two jumps
should be close to the particle length.
\bibitem{wall} 
Although data acquisitions were always taken in the middle of the
sample which is rather thin (about $10 \; \mu m$), the diffusion has
been observed to slow down close to the cell walls.
\bibitem{sm} The interpretation of such behaviors is ambiguous due to the different
temperature dependence of the viscosity parallel and perpendicular
to the director.
\end{references}
\end{document}